\documentclass[12pt,noshowpacs,nofootinbib,notitlepage,amsmath,amssymb]{revtex4-2}
\usepackage{setspace}
\allowdisplaybreaks
\usepackage{graphicx,color}
\usepackage[colorlinks=true,citecolor=blue,linkcolor=blue,urlcolor=blue]{hyperref}
\usepackage[charter]{mathdesign}
\usepackage{multirow}
\usepackage{enumerate}
\usepackage{array}
\usepackage{booktabs}

\DeclareSymbolFontAlphabet{\mathcal}{symbols}
\DeclareSymbolFont{symbols}{OMS}{xmdcmsy}{m}{n}
\DeclareSymbolFont{largesymbols}{OMX}{cmex}{m}{n}
\SetSymbolFont{symbols}{bold}{OMS}{xmdcmsy}{b}{n}

\begin{document}  
\title{\color{blue}\Large Towards rotating 2-2-holes}

\author{Bob Holdom}
\email{bob.holdom@utoronto.ca}
\affiliation{Department of Physics, University of Toronto, Toronto, Ontario, Canada  M5S 1A7}
\begin{abstract}
Static 2-2-hole solutions of quadratic gravity have been investigated to be a possible horizonless replacement for black holes as the endpoint of gravitational collapse. Realistically such objects will form with spin, but rotating 2-2-hole solutions are currently not known. We take some steps here to explore the existence and properties of such solutions. We employ an expansion of the field equations where the expansion parameter is inversely related to the size of the object. This expansion parameter appears explicitly in the trial metrics, and we are able to find solutions of the leading order field equations. These vacuum solutions are candidates to describe most of the interior of a rotating 2-2-hole.
\end{abstract}

\maketitle 

\section{Introduction}
A 2-2-hole \cite{Holdom:2016nek,Holdom:2019ouz,Ren:2019jft} is a horizonless solution of the classical field equations of quadratic gravity, a $2+4$ derivative theory. The interior of a 2-2-hole involves high curvatures and both the two and four derivative terms in the fields equations are needed to describe the interior solution. But 2-2-holes can be macroscopically large ($GM^2\gg1$), and so one would have expected an intrinsic suppression of the four derivative terms because of the smallness of two extra derivatives $\sim (GM)^{-2}$ of the four derivative terms as compared to the factor of $G^{-1}$ in the two derivative terms. This suppression is overcome because the metric itself develops components that become very small in the interior, by a factor of order $\varepsilon\equiv (GM^2)^{-1}$. This factor ends up enhancing the four derivative terms by a relative factor of $\varepsilon^{-1}$, and this is what causes the two sets of terms in the field equations to be similar in magnitude. The interior metric of a static 2-2-hole can be expressed as
\begin{align}
ds^2=-\varepsilon B(r)dt^2+\varepsilon A(r)dr^2 + r^2 d\theta^2 + r^2\sin(\theta)^2 d\phi^2
\label{e1}.\end{align}
We have explicitly factored out the factor of $\varepsilon$, so that $A(r)$ and $B(r)$ are now of order one throughout most of the interior. They both vanish at the origin like $r^2$, a feature that prompted the name ``2-2-hole''.

When $\varepsilon$ is minuscule, to a very good approximation the 2-2-hole solution can be obtained by expanding the field equations to leading order in $\varepsilon$ and solving these somewhat simplified equations for $A(r)$ and $B(r)$. This will describe only the interior. Close to the would-be horizon radius of $2GM$, $\varepsilon A(r)$ and $\varepsilon B(r)$ grow very rapidly, and the curvature invariants drop very rapidly, such as to match onto the Schwarzschild (Schd) solution at a radius just slightly larger than $2GM$. The approximation of dropping the higher order terms breaks down in the transition region and in particular the Schd solution of the exterior requires the full Einstein equations and not just the terms leading in $\varepsilon$.

In this work we shall only consider rotating \textit{vacuum} solutions. In particular we restrict ourselves to the vacuum field equations of the Einstein-Weyl theory,
\begin{align}
m_{\rm Pl}^2(R_{\mu\nu}-\frac{1}{2}g_{\mu\nu}R)-2\frac{m_{\rm Pl}^2}{m_G^2}B_{\mu\nu}=0
.\label{e2}\end{align}
$B_{\mu\nu}=0$ is the traceless Bach tensor, and we take the ghost mass to be $m_G\approx m_{\rm Pl}$. Static 2-2-hole solutions were first obtained as vacuum solutions \cite{Holdom:2002xy} (see also \cite{Lu:2015psa}). More realistic and interesting static solutions that are sourced by a thermal gas were found later \cite{Holdom:2019ouz} (see also \cite{Ren:2019jft}). The associated entropy is easily understood and calculated, and it is somewhat larger than the entropy of a similar size black hole when $m_G\approx m_{\rm Pl}$. This then points to the stability of 2-2-holes. Vacuum solutions, whether static or rotating, are probably not stable,\footnote{Rotating solutions are still stationary and so like the static solutions they are blind to the potential instabilities inherent in four derivative field equations. For time dependent problems, with fluctuations of Planck scale or smaller, input from quantum quadratic gravity, the associated stable quantum field theory, would be needed.} but in the static case the vacuum and non-vacuum solutions are found to be at least qualitatively similar. 

Rotation brings in $\theta$ as well as $r$ dependence and an off-diagonal component of the metric. Correspondingly there is an increased number of components (six) of the field equations that need to be solved. In the static case a hybrid approach based on a series expansion around the origin along with numerical integration proved to be sufficient to study the 2-2-hole solutions, both for the interior and transition regions and the matching to the exterior Schd solution. A rather complete picture of the static thermal 2-2-holes is obtained in \cite{Holdom:2022zzo}. This approach does not seem practical in the rotating case.

We shall exploit the $\varepsilon$ expansion by inserting explicit factors of $\varepsilon$ into trial metrics, and then expanding the field equations in powers of $\varepsilon$. This expansion proves to be quite helpful by reducing the complexity of the leading order components of the field equations. We shall focus on finding the leading order metric, that is the metric that satisfies the different components of the field equations at leading order in $\varepsilon$. Note that the different components may have leading order terms that differ in the power of $\varepsilon$, but all these leading order terms must vanish for a leading order solution. We shall also consider the effects of higher orders in the $\varepsilon$ expansion, and for this we must modify the metric with corrections that are relatively suppressed by $\varepsilon$. But it is only when $r$ approaches the transition region from the inside will the higher order effects be important. The leading order metric should be a good approximation throughout most of the interior.

A neutral rotating 2-2-hole will be characterized by two dimensionful constants, its size $R$ and its spin $a$ (having the same dimension as $R$). Various functions define the various components of the metric and we shall pull out explicit factors of $R$ to be left with dimensionless functions. These are the $F_i(r,\theta)$ with $i=t,r,\theta,\phi,t\phi$ labelling one of the four diagonal components or the one off-diagonal component of the metric. These are actually functions of $r/R$, $a/R$ and $\theta$. Alternatively we may use a dimensionless function $f_i(\theta)$, a function only of $a/R$ and $\theta$.

We are able to make progress with the following two ansatzes for the leading order metric. They differ depending on whether the $g_{tt}$ component is of order $\varepsilon^0$ or $\varepsilon^1$.

Ansatz 1
\begin{align}
ds^2=-f_t(\theta)\,dt^2+\varepsilon F_r(r,\theta)\, dr^2 + R^2 F_\theta(r,\theta) \,d\theta^2 + R^2f_\phi(\theta)\, d\phi^2\pm 2 R f_{t\phi}(\theta)\, dt\, d\phi
,\label{e4}\end{align}

Ansatz 2
\begin{align}
ds^2=-\varepsilon F_t(r,\theta)\,dt^2+\varepsilon F_r(r,\theta)\, dr^2 + R^2F_\theta(r,\theta) \,d\theta^2 + R^2F_\phi(r,\theta)\, d\phi^2\pm 2R F_{t\phi}(r,\theta)\, dt\, d\phi
.\label{e16}\end{align}
All functions are generally non-negative, except possibly $f_t(\theta)$ and $F_t(r,\theta)$. The latter functions may describe ergoregions when negative. Ansatz 2 involves more general functions of $r$ and $\theta$, and with both $g_{tt}$ and $g_{rr}$ of order $\varepsilon$ it is more in line with the static case. But it is still interesting to compare and contrast ansatz 1 with ansatz 2, and in the next two sections we consider these two ansatzes in turn. In each case we will be able to solve the leading order equations and thus obtain constraints on the leading order metric. The leading order metric is not determined completely, and we expect that any freedom left in it will be fixed in a full solution that matches the interior solution to the exterior solution. This might only be possible for one of the two ansatzes.
 
We shall also consider corrections to these leading order metrics as defined by the following corrected line element,
\begin{align}
ds^2_{\rm corrected}=ds^2&+\varepsilon^p G_t(r,\theta)\,dt^2+\varepsilon^2 G_r(r,\theta)\, dr^2+ \varepsilon G_\theta(r,\theta) \,d\theta^2 \nonumber\\& + \varepsilon G_\phi(r,\theta)\, d\phi^2+\varepsilon G_{t\phi}(r,\theta)\, dt\, d\phi
\label{e15},\end{align}
where $p=1,2$ for Ansatz 1 and 2 respectively.  It will turn out to be useful to know at what order the various components of the field equations depend on the correction terms in the metric.

For both ansatzes we shall find a Ricci-flat metric at leading order in $\varepsilon$. By this we mean that $R_{\mu\nu}$ is reduced in size by one order of $\varepsilon$ after the leading order equations are solved. This result implies that these solutions are in no way smoothly related to the static 2-2-hole solutions, which are certainly not Ricci-flat at leading order in $\varepsilon$. Our results for the rotating case are therefore pointing towards a new class of solutions. We view the Ricci-flatness at leading order as accidental, and we do not expect it to extend to next-to-leading order as well. If it held to all orders then we should get back the unique rotating vacuum solution of the Einstein equations, which is the Kerr metric. But our leading-order solutions are certainly not Kerr-like.

One might wonder about other expectations for how the interior metric of a 2-2-hole should depend on $a$ and $\theta$. In this connection it is useful to compare the Kerr metric (in Boyer-Lindquist coordinates) in the limit $r\to\infty$ to the less familiar limit $r\to0$,
\begin{align}
ds^2_{\rm Kerr}&\underset{r\to \infty}{=}-\,dt^2+dr^2 + r^2 \,d\theta^2 + r^2\sin(\theta)^2\, d\phi^2\pm \frac{a}{r}4 GM \sin \left(\theta \right)^{2}\, dt\, d\phi\\
&\underset{r\to0}{=}-\,dt^2+\cos(\theta)^2\, dr^2 + a^2\cos(\theta)^2 \,d\theta^2 + a^2\sin(\theta)^2\, d\phi^2\pm \frac{r}{a}\frac{4 GM \sin \left(\theta \right)^{2}}{\cos \left(\theta \right)^{2}}\, dt\, d\phi
.\nonumber\end{align}
They are very different, and so we should be cautious about preconceived notions for the $a$ and $\theta$ dependence inside a rotating 2-2-hole.

The main results for the two ansatzes are described in Sections \ref{s2} and \ref{s3}. We also discuss the propagation of waves and particles on these spacetimes. In Section \ref{s4} we briefly describe an attempt to find a rotating 2-2-hole solution that would be smoothly connected to the static metric solution when $a\to0$. Some conclusions are given in Section \ref{s5}.

\section{Anzatz 1}\label{s2}
With the metric in (\ref{e4}), the Einstein and Bach tensors in the field equations start at order $\varepsilon^{-1}$ and $\varepsilon^{-2}$ respectively. As we have explained, this allows for the actual relative size of the Einstein and Bach tensors to be comparable in the interior of macroscopically large objects. The Einstein tensor has three non-vanishing components at this order, the $tt$, $\phi\phi$ and $t\phi$ components. The $\theta\theta$ component accidentally vanishes; accidental zeros arise for this ansatz because of the lack of $r$ dependence in $g_{tt}$, $g_{\phi\phi}$ and $g_{t\phi}$. All three nonzero components of the Einstein tensor turn out to be proportional to
\begin{align}
F_r \! \left(r , \theta \right) F_\theta'\! \left(r , \theta \right)^{2}+F_\theta\! \left(r , \theta \right) F_\theta'\! \left(r , \theta \right) F_r' \! \left(r , \theta \right)-2 F_r \! \left(r , \theta \right) F_\theta\! \left(r , \theta \right) F_\theta''\! \left(r , \theta \right)
.\end{align}
(Our notation is such that a prime always denotes a derivative with respect to the first argument of the function.) This vanishes when
\begin{align}
F_r(r,\theta)=\frac{f_r(\theta)}{F_\theta(r , \theta )} R^2 F_\theta'(r , \theta )^{2}
\label{e3}.\end{align}
$f_r(\theta)\geq0$ is another dimensionless function of $a/R$ and $\theta$.

The Bach tensor has all four of the $tt$, $\phi\phi$, $t\phi$ and $\theta\theta$ components non-vanishing at this order. Each is a complicated function involving up to four derivatives of $F_\theta(r,\theta)$ and three derivatives of $F_r(r,\theta)$, but we have verified that these four components of the Bach tensor also vanish when (\ref{e3}) is satisfied. Thus at this leading order the field equations can be satisfied by having the Einstein and Bach terms separately vanishing, with the simple relationship between $F_r(r,\theta)$ and $F_\theta(r,\theta)$ in (\ref{e3}).

With relation (\ref{e3}) in place, we can now inspect the field equations at next order, that is at $\varepsilon^{0}$ and $\varepsilon^{-1}$ for the Einstein and Bach tensors respectively. At this order the field equations will have dependence on the next order corrections in the metric, as in (\ref{e15}). The $tt$, $\phi\phi$, $t\phi$ and $\theta\theta$ components of the Einstein and Bach tensors are nonzero and we have verified that each of these components are sensitive to at least some or all of the various $G_i(r,\theta)$'s and their derivatives. The expressions are lengthy and complicated, especially for the Bach tensor. But the many undetermined functions leaves open the possibility that the field equations at next order can be satisfied. In this case we expect that this will occur through cancellations between the Einstein and Bach terms that are separately non-vanishing.

There is one additional component that becomes nonzero at this order, and this is the $r\theta$ component of the Einstein tensor. (This component of the Bach tensor is accidentally zero.) This nonzero Einstein component is receiving its leading contribution at this order, and so it is not sensitive to the $G_i(r,\theta)$'s. It therefore implies another constraint on the leading order metric. This component is proportional to
\begin{align}
f_\phi(\theta)f_t'(\theta)+f_t(\theta)f_\phi'(\theta)+2f_{t\phi}(\theta)f_{t\phi}'(\theta)
,\end{align}
which vanishes when 
\begin{align}
f_t(\theta)f_\phi(\theta)+f_{t\phi}(\theta)^2=C
,\label{e5}\end{align}
for some dimensionless constant $C$. Note that $C$ can also depend on $a/R$.

If we expand to one further order (orders $\varepsilon^1$ and $\varepsilon^0$ respectively) then the $r r$ component finally shows up. But this component at this order is sensitive to the corrected metric, meaning that it was accidentally zero at the previous order. It therefore provides no further constraint on the leading order metric.

\subsection{Invariants}

With the results in (\ref{e3}) and (\ref{e5}) we find that the volume element is
\begin{align}
\sqrt{-g}&=\left(\varepsilon C R^6 f_r(\theta)F_\theta'(r,\theta)^2\right)^\frac{1}{2}
.\label{e6}\end{align}
We see a $\sqrt{\varepsilon}$ suppression of the volume element, to be compared with a $\varepsilon$ suppression for the static 2-2-hole. We have already seen that $f_r(\theta)\geq0$ and thus we must have $C>0$. $f_t(\theta)$ is the only function that could be negative, and we see that (\ref{e5}) with $C>0$ still permits $f_t(\theta)$ to have either sign. When negative then $g_{tt}\geq0$ would correspond to an ergoregion that extends to the origin of the 2-2-hole. 

Let us now consider the quadratic curvature invariants. We find that two are enhanced by $\varepsilon^{-1}$,
\begin{align}
&R_{\mu\nu\rho\sigma}R^{\mu\nu\rho\sigma}=C_{\mu\nu\rho\sigma}C^{\mu\nu\rho\sigma}=\nonumber\\&\frac{1}{\varepsilon}\frac{\left((C -f_{t\phi} (\theta)^{2}) f_t'(\theta)^{2}+2 f_{t\phi}' (\theta) f_{t\phi} (\theta) f_t' (\theta) f_t (\theta)-f_{t\phi}' (\theta)^{2} f_t (\theta)^{2}\right)}{R^4 C f_t (\theta)^{2} f_r (\theta) F_\theta\! \left(r , \theta \right)^{2}}
\label{e7}\end{align}
We have used (\ref{e5}) to eliminate $f_\phi(\theta)$. This result is stable under corrections to the leading order metric, that is it is independent of the $G_i(r,\theta)$'s in (\ref{e15}). $R_{\mu\nu}R^{\mu\nu}$, like $R$, turns out to be of order $\varepsilon^0$, and it does depend on the $G_i(r,\theta)$'s. From (\ref{e7}) we notice that $f_t(\theta)$ and $f_r(\theta)$ must not vanish to avoid a singularity at all $r$.

\subsection{Propagation}

We first consider the wave equation for a massless scalar field $\varphi(r,\theta,\phi,t)$ on this 2-2-hole background, $\Box\varphi=0$. At leading order in $\varepsilon$ only the $t$- and $r$-derivative terms remain. The neglect of the angular derivative terms at this stage assumes that the angular momenta are not so large as to compensate for the $\varepsilon$ suppression. The single $r$-derivative terms can be eliminated if we define
\begin{align}
\varphi(r,\theta,\phi,t)=e^{-i\omega t}\sqrt {{\frac {F_\theta' \left( r,\theta \right) }{F_\theta \left( r,\theta\right) }}}\Phi(r,\theta,\phi)
.\end{align}
Then the leading order wave equation takes the form
\begin{align}
\left({\frac {\partial ^{2}}{\partial {r}^{2}}}+\omega^2-V(r,\theta)\right)\Phi \left( r,\theta,\phi \right)=0
,\label{e19}\end{align}
where the potential is
\begin{align}
V(r,\theta)={\frac { 2 F_\theta \left( r,
\theta \right) ^{2}  F_\theta' \left( r,\theta \right)
  F_\theta''' \left( r,\theta \right) -3 F_\theta \left( r,\theta \right) ^{2}  F_\theta'' \left( r,\theta \right)   ^{2} + F_\theta' \left( r,\theta \right) ^{4} }{ 4 F_\theta \left( r,\theta
 \right) ^{2} F_\theta' \left( r,\theta \right) ^{2}}}
.\label{e20}\end{align}
We see that both the wave equation and the curvatures can be singular at the origin if $F(r,\theta)$ vanishes there. It is also possible for the potential to be singular while the curvatures are not, when $F'(r,\theta)$ but not $F(r,\theta)$ vanishes at the origin; for example if $F'(r,\theta)\sim r$ then $V(r,\theta)\sim r^{-2}$ with a negative residue. The static 2-2-hole is quite different, there the curvatures are singular while the wave equation, including the angular part, is regular at the origin.

Particle-like motion is described by the geodesic equations, corresponding to a small wave-length (eikonal) limit of the wave equations. This limit is essentially the opposite to that used to obtain the form of the wave equation above. The geodesic equations for a massless particle describes a path $(t(\zeta),r(\zeta),\theta,\phi(\zeta))$ as a function of the affine parameter $\zeta$, where for simplicity we have set $\theta$ to a constant. We then have
\begin{align}
\frac{d^2t(\zeta)}{d\zeta^2}=0\quad\frac{d^2\phi(\zeta)}{d\zeta^2}=0
,\end{align}
and a radial equation that is solved by
\begin{align}
\sqrt{F_\theta(r(\zeta),\theta)}=c_0(\theta)+c_1(\theta)\zeta
.\end{align}

The existence and stability of possible light ring orbits can be found via the formalism discussed in \cite{Cunha:2017qtt}. Two potentials are defined by
\begin{align}
H_\pm(r,\theta)=\frac{-g_{t\phi}\pm\sqrt{g_{t\phi}^2-g_{tt}g_{\phi\phi}}}{g_{\phi\phi}}
.\label{e21}\end{align}
The existence of a light ring corresponds to a stationary point occurring for either potential, and a stable light ring corresponds to this being a local minimum of that potential. Stable light rings are thought to imply a spacetime instability that may destroy the object.\footnote{Realistic 2-2-holes are sourced by a hot thermal gas, and the effect of light rings occurring within such a medium may need to be considered further.} The metric components in (\ref{e21}) precisely those that are $r$-independent in anzatz 1, and so these potentials are independent of $r$,
\begin{align}
H_\pm(r,\theta)=\frac{f_t(\theta)}{(\sqrt{C}\mp f_\phi(\theta))R}
.\end{align}
This degeneracy is expected to be broken at next order in $\varepsilon$, and so at this stage we can say little about interior light rings. Ansatz 2 will give $r$-dependent potentials at leading order.

\subsection{A particular choice}

$F_\theta(r,\theta)$ is the remaining undetermined $r$-dependent function in the metric of ansatz 1. Of all the metric components, $g_{\theta\theta}$ is the only one that could possibly be the same as the Kerr metric. The static 2-2-hole solution already strongly modifies $g_{tt}$ and $g_{rr}$ from the Schd metric, and then rotation implies a non-vanishing $g_{t\phi}$ that points to a modification of $g_{\phi\phi}$ as well. This leaves $g_{\theta\theta}$, and so for ansatz 1 we consider the Kerr choice $g_{\theta\theta}=R^2F_\theta(r,\theta)=r^2+a^2\cos(\theta)^2$. Then the curvature invariants in (\ref{e7}) are singular only when both $r=0$ and $\theta=\pi/2$, as with the Kerr metric. They also show that the leading order metric is asymptotically flat, although this is not relevant for its role as an approximate interior metric. The metric in this case is
\begin{align}
ds^2=&-f_t(\theta)\,dt^2+\varepsilon\frac{4f_r(\theta)r^2}{r^2+a^2\cos(\theta)^2}\, dr^2 + (r^2+a^2\cos(\theta)^2) \,d\theta^2 \nonumber\\&+ R^2\frac{C-f_{t\phi}(\theta)^2}{f_t(\theta)}\, d\phi^2\pm 2 R f_{t\phi}(\theta)\, dt\, d\phi
.\end{align}
The various $f_i(\theta)$ and $C$ can depend on $a/R$. We see $g_{rr}\sim r^2$, at least for small $r/a$, which is similar to the $g_{rr}\sim r^2$ behaviour of a static 2-2-hole. In addition
\begin{align}
\sqrt{-g}&=2Rr\left(\varepsilon C f_r(\theta)\right)^\frac{1}{2}
.\end{align}

\section{Ansatz 2}\label{s3}
This ansatz for the leading order metric is given in (\ref{e16}). Once again the leading order terms of the Einstein and Bach tensors are of order $\varepsilon^{-1}$ and $\varepsilon^{-2}$ respectively, making them of similar size. At this order, and for each of the two tensors, the three non-vanishing components are $\theta\theta$, $\phi\phi$ and $\phi t$. Looking first at the Einstein tensor, the $\theta\theta$ component is proportional to
\begin{align}
&4 F_{t\phi}''\! \left(r , \theta \right) F_r \! \left(r , \theta \right) F_{t\phi}\! \left(r , \theta \right)-F_{t\phi}'\! \left(r , \theta \right)^{2} F_r \! \left(r , \theta \right)-2 F_{t\phi}'\! \left(r , \theta \right) F_r' \! \left(r , \theta \right) F_{t\phi}\! \left(r , \theta \right)
.\end{align}
This vanishes when
\begin{align}
F_r(r,\theta)= \frac{f_r(\theta)}{\sqrt{F_{t\phi}(r , \theta )}}R^2 F_{t\phi}'(r , \theta )^{2}
.\label{e10}\end{align}
When this result is inserted into the $\phi t$ component, the latter becomes proportional to
\begin{align}
&F_r\! \left(r , \theta \right)^{2} F_{t\phi}'\! \left(r , \theta \right)^{3}-4 F_r\! \left(r , \theta \right) F_{t\phi}\! \left(r , \theta \right)^{2} F_{t\phi}'\! \left(r , \theta \right) F_r''\! \left(r , \theta \right)
\nonumber\\&+4 F_r\! \left(r , \theta \right) F_{t\phi}\! \left(r , \theta \right)^{2} F_{t\phi}''\! \left(r , \theta \right) F_r'\! \left(r , \theta \right)\label{e14}\\\nonumber&-3 F_r\! \left(r , \theta \right) F_{t\phi}\! \left(r , \theta \right) F_{t\phi}'\! \left(r , \theta \right)^{2} F_r'\! \left(r , \theta \right)+2 F_{t\phi}\! \left(r , \theta \right)^{2} F_{t\phi}'\! \left(r , \theta \right) F_r'\! \left(r , \theta \right)^{2}
.\end{align}
This vanishes when
\begin{align}
F_\theta(r,\theta)=\frac{f_\theta(\theta)}{\sqrt{F_{t\phi}(r , \theta )}}
.\label{e11}\end{align}
When the last two relations are inserted into the $\phi \phi$ component, the latter becomes proportional to
\begin{align}
&F_\phi\! \left(r , \theta \right) F_{t\phi}'\! \left(r , \theta \right)^{3}+F_{t\phi}\! \left(r , \theta \right)^{2} F_{t\phi}'\! \left(r , \theta \right) F_\phi''\! \left(r , \theta \right)\nonumber\\&-F_{t\phi}\! \left(r , \theta \right)^{2} F_\phi'\! \left(r , \theta \right) F_{t\phi}''\! \left(r , \theta \right)-F_{t\phi}\! \left(r , \theta \right) F_{t\phi}'\! \left(r , \theta \right)^{2} F_\phi'\! \left(r , \theta \right).\label{e13}\end{align}
This vanishes when
\begin{align}
F_\phi(r,\theta)= F_{t\phi}(r , \theta )\left(f_\phi(\theta)+\hat f_\phi(\theta)\ln(F_{t\phi}(r,\theta))\right)
.\label{e12}\end{align}
Now when we turn to the Bach tensor, we find that these three relations also cause the vanishing of the order $\varepsilon^{-2}$ contributions to the $\theta\theta$, $\phi\phi$ and $\phi t$ components.\footnote{Note that (\ref{e14}) has a more general solution than that given in (\ref{e11}), but then the three Bach tensor components do not all vanish.} So once again the Einstein and Bach tensors vanish independently, something that is not expected to hold at next order in the $\varepsilon$ expansion.

With these relations in place then components start at order $\varepsilon^{0}$ and $\varepsilon^{-1}$ for the Einstein and Bach tensors respectively. We have verified that the metric correction terms in (\ref{e15}) contribute to the $\theta\theta$, $\phi\phi$ and $\phi t$ components at this order. In principle these field equations at this order would help to determine the $G_i(r,\theta)$ functions. We are more interested in the $t t$, $r \theta$ and $r r$ components of both tensors, since these components first appear at this order and so these do not receive corrections. The $r\theta$ component of the Bach tensor accidentally vanishes and $r r$ component of both tensors happen to vanish due to the relations we have already applied. The required vanishing of the $t t$ and $r \theta$ components must be due to further constraints on the leading order metric. We find that the $t t$ component of the Einstein tensor is proportional to the same expression as in (\ref{e13}), but with $F_\phi(r,\theta)$ replaced by $F_t(r,\theta)$. Thus this vanishes when
\begin{align}
F_t(r,\theta)= F_{t\phi}(r , \theta )\left(f_t(\theta)+\hat f_t(\theta)\ln(F_{t\phi}(r,\theta))\right)
.\label{e17}\end{align}
This also causes the $tt$ component of the Bach tensor to vanish. Lastly, the $r \theta$ component of the Einstein tensor is proportional to $f_r'(\theta)$, and thus we must replace $f_r(\theta)$ by a constant $C$ that can still depend on $a/R$.

We can expand the Einstein and Bach tensors to one further order, that is to order $\varepsilon^1$ and $\varepsilon^0$ respectively. The only nonzero components for either tensor are $t t$, $r r$ and $r \theta$, but these components have already appeared at lower order and so they will receive corrections from the corrected metric. Thus we do not obtain any further constraints on the leading order metric. From the constraints that we do have, all the $r$-dependent functions have been expressed in terms of $F_{t\phi}(r,\theta)$.

\subsection{Invariants}
The volume element at leading order in $\varepsilon$ is
\begin{align}
\sqrt{-g}&=\left(\varepsilon R^4 F_r(r,\theta) F_\theta(r,\theta) F_{t\phi}(r,\theta)^2\right)^\frac{1}{2}\nonumber\\
&=\left(\varepsilon C R^6  f_\theta(\theta) F_{t\phi}(r,\theta) F_{t\phi}'(r,\theta)^2\right)^\frac{1}{2}
.\label{e8}\end{align}
Again we have a $\sqrt{\varepsilon}$ dependence. The dependence on $F_t(r,\theta)$ only occurs at higher order, and this is interesting because it allows $F_t(r,\theta)$ to be negative. In fact from (\ref{e17}), appropriate $f_t(\theta)$ and $\hat f_t(\theta)$ can give rise to an interior surface of some shape where $F_t(r,\theta)$ changes sign. An ergoregion could lie inside or outside this surface.

The curvature invariants take a simple form
\begin{align}
&R_{\mu\nu\rho\sigma}R^{\mu\nu\rho\sigma}=C_{\mu\nu\rho\sigma}C^{\mu\nu\rho\sigma}=\frac{1}{\varepsilon^2}\frac{3}{4C^2R^4F_{t\phi}(r,\theta)^3}
\label{e9}.\end{align}
Again this is not sensitive to higher order corrections to the metric. ($R_{\mu\nu}R^{\mu\nu}$ on the other hand, like $R$, turns out to be of order $\varepsilon^0$ and is sensitive to corrections.) Notice also that these invariants are proportional to $\varepsilon^{-2}\sim R^4/\ell_{\rm Pl}^4$, which means that they are now of order one in Planck units. As compared to ansatz 1, where the curvature invariants are proportional to $\varepsilon^{-1}$, ansatz 2 is thus more similar to the situation for a static 2-2-hole.

\subsection{Propagation}
The scalar wave equation $\Box\varphi=0$ at leading order in $\varepsilon$ yields a very similar result to ansatz 1. If we define
\begin{align}
\varphi(r,\theta,\phi,t)=e^{-i\omega t}\sqrt {{\frac {F_{t\phi}' \left( r,\theta \right) }{F_{t\phi} \left( r,\theta\right) }}}\Phi(r,\theta,\phi)
,\end{align}
then we get the same wave equation as in (\ref{e19}) and (\ref{e20}), only with $F_\theta(r,\theta)$ replaced by $F_{t\phi}(r,\theta)$. With this change the previous discussion applies, and in particular the wave equation potential can become singular when either $F_{t\phi}(r,\theta)$ or $F'_{t\phi}(r,\theta)$ vanishes.

The geodesic equations are complicated and so to permit a solution we set $\hat f_t(\theta)=\hat f_\phi(\theta)=0$. In this case, for the same path $(t(\zeta),r(\zeta),\theta,\phi(\zeta))$ as before, we find
\begin{align}
\frac{dt(\zeta)}{d\zeta}=-\frac{c_1 f_\phi(\theta)}{2F_{t\phi}(r(\zeta),\theta)}    \quad\frac{d\phi(\zeta)}{d\zeta}=-\frac{c_1}{RF_{t\phi}(r(\zeta),\theta)}
.\end{align}
This result has also been constrained by the leading order part of the radial equation. The next-to-leading part of the radial equation is then solved by
\begin{align}
\frac{d}{d \zeta}r \! \left(\zeta \right) =
\frac{\sqrt{C F_{t\phi} \! \left(r \! \left(\zeta \right), \theta \right)^{\frac{3}{2}} \left(c_2 F_{t\phi} \! \left(r \! \left(\zeta \right), \theta \right) +f_\phi \! \left(\theta \right)^{2} f_t \! \left(\theta \right)\right)}}{2 C F_{t\phi} \! \left(r \! \left(\zeta \right), \theta \right) F_{t\phi}' \! \left(r \! \left(\zeta \right), \theta \right) R}
.\end{align}

The potentials defined in (\ref{e21}) for the determination of light rings are 
\begin{align}
H_+(r,\theta)&=\frac{\varepsilon}{2R}\left(f_t(\theta)+\hat f_t(\theta)\ln(F_{t\phi}(r,\theta))\right),\\
H_-(r,\theta)&=-\frac{2}{R}\frac{1}{f_\phi(\theta)+\hat f_\phi(\theta)\ln(F_{t\phi}(r,\theta))}-\frac{\varepsilon}{2R}\left(f_t(\theta)+\hat f_t(\theta)\ln(F_{t\phi}(r,\theta))\right)
.\end{align}
These potentials are of a form that can easily avoid any stationary points. Notice that the denominator of the first term of $H_-(r,\theta)$ is proportional to $g_{\phi\phi}\geq0$, and that one way to meet this constraint is to set $\hat f_\phi(\theta)=0$. This is why the next order term is also shown.

\subsection{A particular choice}
The results for the volume element and the curvature invariants in (\ref{e8}) and (\ref{e9}), as well as the wave equation, are showing that $F_{t\phi}(r,\theta)$ for ansatz 2 is playing a similar role to $F_\theta(r,\theta)$ for ansatz 1. Thus it helps to compare the two ansatzes if the particular choice we made for $F_\theta(r,\theta)$ is carried over to $F_{t\phi}(r,\theta)$ for ansatz 2, that is $R^2F_{t\phi}(r,\theta)= r^2+a^2\cos(\theta)^2\equiv\Sigma$. The metric becomes
\begin{align}
ds^2=&-\varepsilon\frac{\Sigma}{R^2}\left(f_t(\theta)+\hat f_t(\theta)\ln(\frac{\Sigma}{R^2})\right)\,dt^2+\varepsilon\frac{4Cr^2}{R\sqrt{\Sigma}}\, dr^2 \nonumber\\&+ \frac{R^3f_\theta(\theta)}{\sqrt{\Sigma}} \,d\theta^2 + \Sigma f_\phi(\theta)\, d\phi^2\pm \frac{2 \Sigma}{R}\, dt\, d\phi
.\end{align}
The various $f_i(\theta)$ and $C$ can still depend on $a/R$. We have set $\hat f_\phi(\theta)=0$ to ensure that $g_{\phi\phi}\geq0$, since $\Sigma$ ranges from 0 to $\infty$. We again see $g_{rr}\sim r^2$ at small $r/a$. In addition
\begin{align}
\sqrt{-g}&=2\sqrt{C\varepsilon f_\theta(\theta)\Sigma r^2}
,\label{e22}\end{align}
which, other than the $\sqrt{\varepsilon}$, is rather similar to the Kerr metric where $\sqrt{-g}=\sin(\theta)\Sigma$. The curvature invariants are simply
\begin{align}
&R_{\mu\nu\rho\sigma}R^{\mu\nu\rho\sigma}=C_{\mu\nu\rho\sigma}C^{\mu\nu\rho\sigma}=\frac{1}{\varepsilon^2}\frac{3}{4}\frac{R^2}{C^2\Sigma^3}
.\label{e23}\end{align}
These invariants for the Kerr metric also have a singularity at $\Sigma=0$ and the same $1/r^6$ behaviour at large $r$. The main difference is the $\varepsilon^{-2}$ enhancement for the 2-2-hole interior.

Finally we give an example of how the explicit $a$ dependence in the metric can change without changing the results in (\ref{e22}) and (\ref{e23}). We could choose $F_{t\phi}(r,\theta)=\Sigma/(a R)$ and then redefine $(f_\theta(\theta),C)\to(\frac{a}{R})^\frac{3}{2} (f_\theta(\theta),C)$ and $(f_\phi(\theta),f_t(\theta),\hat f_t(\theta))\to \frac{a}{R} (f_\phi(\theta),f_t(\theta),\hat f_t(\theta))$. In this case the metric is
\begin{align}
ds^2=&-\varepsilon\frac{\Sigma}{R^2}\left(f_t(\theta)+\hat f_t(\theta)\ln(\frac{\Sigma}{a R})\right)\,dt^2+\varepsilon\frac{4Cr^2}{R\sqrt{\Sigma}}\, dr^2 \nonumber\\&+ \frac{a^2R f_\theta(\theta)}{\sqrt{\Sigma}} \,d\theta^2 + \Sigma f_\phi(\theta)\, d\phi^2\pm \frac{2 \Sigma}{a}\, dt\, d\phi
.\end{align}

\section{An attempt with a more direct approach}\label{s4}
A recent new representation of the Kerr metric has appeared \cite{Baines:2022xnh}, and it leads to an obvious starting point in the search for a rotating 2-2-hole solution. Using their notation, these authors find that
\begin{align}
(g_{\rm Kerr})_{ab}=g_{AB}{e^A}_a{e^B}_b
,\end{align}
where the ``co-tetrad'' is
\begin{align}
{e^A}_a=
\left[\begin{array}{cccc}
\sqrt{\frac{a^{2}+x^{2}}{x^{2}+a^{2} \cos \left(\theta \right)^{2}}} & 0 & 0 & -\frac{\sqrt{a^{2}+x^{2}}\, a \sin \left(\theta \right)^{2}}{\sqrt{x^{2}+a^{2} \cos \left(\theta \right)^{2}}}
\\
 0 & \sqrt{\frac{x^{2}+a^{2} \cos \left(\theta \right)^{2}}{a^{2}+x^{2}}} & 0 & 0
\\
 0 & 0 & \sqrt{x^{2}+a^{2} \cos \! \left(\theta \right)^{2}} & 0
\\
 -\frac{a \sin \left(\theta \right)}{\sqrt{x^{2}+a^{2} \cos \left(\theta \right)^{2}}} & 0 & 0 & \frac{\left(a^{2}+x^{2}\right) \sin \left(\theta \right)}{\sqrt{x^{2}+a^{2} \cos \left(\theta \right)^{2}}}
 \end{array}\right]
,\end{align}
and the ``tetrad-component metric'' is
\begin{align}
g_{AB}=
\left[\begin{array}{cccc}
-f(r) & 0 & 0 & 0
\\
 0 & \displaystyle\frac{1}{f(r)} & 0 & 0
\\
 0 & 0 & 1 & 0
\\
0 & 0 & 0 & 1
 \end{array}\right], \quad f(r)=1-\frac{2mr}{r^2+a^2}
.\end{align}
The essence of the Kerr metric is contained in $g_{AB}$, since replacing $g_{AB}$ by $\eta_{AB}$ would yield the metric for flat space in oblate spheroidal coordinates.

This suggests that in the same way a static 2-2-hole is obtained from the Schd metric with the replacements $-g_{tt}=1-2m/r\to\varepsilon B(r)$ and $g_{rr}=1/(1-2m/r)\to\varepsilon A(r)$, we should make the same replacements in the tetrad-component metric $g_{AB}$, that is $-g_{tt}=f(r)\to\varepsilon B(r)$ and $g_{rr}=1/f(r)\to\varepsilon A(r)$. With this ansatz the dimensionless functions $A(r)$ and $B(r)$ are functions of $r/R$ and $a/R$, but not $\theta$. This construction is guaranteed to produce a metric that goes smoothly to the form of the static 2-2-hole metric in (\ref{e1}) when $a\to0$. A point of interest for our discussion in section \ref{s2} is that the metric from this ansatz has $g_{\theta\theta}=r^2+a^2\cos(\theta)^2$.

With this metric we have expanded the field equations in (\ref{e2}) to leading order in $\varepsilon$. For $a\neq0$ the leading order terms of the Einstein and Bach tensors are now of order $\varepsilon^{-2}$ and $\varepsilon^{-4}$ respectively. Derivatives of $A(r)$ and $B(r)$ happen not to appear at leading order, and this makes it quite easy to see that no solution of the vacuum equations for $A(r)$ and $B(r)$ is possible. Some attempt to use a more general form of $g_{AB}$ was also  not successful. So thus far we have found no evidence for a rotating solution that smoothly connects to the static solution.

\section{Conclusion}\label{s5}
We have utilized what is effectively a large size or large mass expansion to approach the problem of describing the interior of a rotating 2-2-hole. The expansion parameter is a very small quantity $\varepsilon\equiv (GM^2)^{-1}$ that appears explicitly in the metric. This causes the four derivative terms in the field equations to be of the same order as the two derivative terms.

We have solved the leading order field equations for two ansatzes. Ansatz 2 produces a result that is qualitatively more similar to the static 2-2-hole, since like the later, two quadratic curvature invariants end up enhanced by a factor of $\varepsilon^{-2}$. This is what produces Planckian curvatures over a macroscopically large region. On the other hand the spacetime for both anzatzes is Ricci-flat up to corrections that are next order in $\varepsilon$. This is not the case for the static 2-2-hole solution.

For each of the two rotating ansatzes we displayed a particular choice of the remaining $r$-dependent function that in turn gives a curvature singularity at $r^2+a^2\cos(\theta)^2=0$. We note that the static solution exhibits a curvature singularity at $r=0$ that is benign with respect to the wave equation. In fact this singularity is associated with the fact that there are no stable light rings in the static 2-2-hole spacetime. For the rotating case it also appears that stable light rings can be avoided, but we cannot fully answer this question without having the full matching to the exterior metric.

We have not proven that one of our solutions provides the true leading order description of the rotating 2-2-hole interior. A proof would entail demonstrating a successful matching to the external Kerr solution. On the other hand there should be \textit{some} leading order description, and at this stage it is difficult to see how other analytical solutions could be found. If we are on the way to the true description, then this would imply that the known static 2-2-hole solution may be an isolated solution that only applies when $a$ is exactly zero. This may not be a physically realizable situation, and so it could be that one should rather take the $a\to0$ limit of the rotating solutions to approach the static case.

If it turns out that we are not yet on the right track to find the rotating solutions, then it would seem to require a more clever approach to generalize the static case, and in particular an approach more clever than the one presented in Section \ref{s4}.

%\acknowledgments

\end{document}